\documentstyle[prl,aps]{revtex}
\begin{document}
\preprint{LPTENS-97/18}
\twocolumn[\hsize\textwidth\columnwidth\hsize\csname@twocolumnfalse\endcsname
\title{Fluctuation-Dissipation theorems and entropy production in relaxational systems }
\author{Leticia F. Cugliandolo{$^*$}\cite{add1}, David S. Dean{$^{*}$}\cite{add2} 
and Jorge Kurchan{$^{**}$}\cite{add3}}
\address{ 
$^{*}$
CNRS-Laboratoire de Physique Th\'{e}orique de l'Ecole Normale 
Sup\'{e}rieure\cite{add2} \\
24, rue Lhomond, F-75231 Paris Cedex 05, France
\\
$^{**}$\'Ecole Normale Sup\'erieure de Lyon, 
46, All\'ee d'Italie, F-69364 Lyon Cedex 07, France
}
\maketitle
\begin{abstract}

We show  that for stochastic dynamical systems out of equilibrium 
the violation of the fluctuation-dissipation equality is bounded 
by  a function of the entropy production.
The result applies to
a  much wider situation than  `near equilibrium', 
comprising diffusion as well as  glasses and other macroscopic systems 
far from equilibrium.
For aging systems this  bounds the age-frequency regimes in which
the susceptibilities 
satisfy FDT  in terms of the rate of
decay of the  $H$-function, 
a question  intimately related to the reading 
of a thermometer placed in contact with the  system.

$\;$ \newline

\end{abstract}

\twocolumn 
\vskip.5pc]
\narrowtext

Many systems encountered in Nature have slow dynamics
but are not `near equilibrium' in any obvious sense.
Typical examples are
(i) Glassy systems long time after the preparation procedure  has finished,
{\em e.g.} long after a temperature quench.
(ii) Domain growth and phase separation.
(iii) Systems  that can be kept `far' from 
equilibrium  by a {\em small} external power input   done by
stationary non-conservative forces and/or 
periodically time-dependent forces.
(iv) Diffusion in non-compact spaces.

A slow, persistent, out of equilibrium regime
is possible because 
in the thermodynamic limit the ergodic time 
is very large compared to the experimental time scales 
--- this is the case of the first three examples --- 
or due to the 
abscence of confining potentials in the case of diffusion.
Indeed,  in contrast with  the usually 
studied  situation \cite{Kubo,Mazur},
none of the cases we mentioned can be 
viewed as a small perturbation about equilibrium
such that the systems would return quickly to equilibrium as soon as
the perturbation is removed.

Moreover,  these systems may
not be regarded in general as `metastable'.
Specifically, one cannot consider a glass or a system undergoing 
domain growth to be completely 
equilibrated within a  fixed sector of phase-space, 
as in the case of {\em e.g.} diamond, because the  probability
distribution in phase-space changes continually. Indeed, 
by measuring two-time correlation or response functions
one can at any time determine the age of a glass
 \cite{St,Cuku,review,numerics}.

Bearing this in mind, a relevant question is whether some specifically 
equilibrium properties are also present in these systems.
In this paper we concentrate on the study of the 
fluctuation-dissipation theorem
(of the `first kind', FDT \cite{Kubo}). 
A physical interpretation of this FDT is that
its validity is a necessary condition 
for a thermometer in contact only with the system to
register the bath's temperature \cite{Cukupe}.

Working with dissipative dynamics, we show that 
the violations of the
fluctuation-dissipation equalities 
vanish with a quantity that can be identified  
with the entropy production rate.

This leads us to suggest that there are  unifying features
 in  situations with  small  entropy 
production rate, whatever its origin
(aging, work done by non-conservative and/or time-dependent forces, temperature
gradients, etc). 
 This comprises a much wider class of systems than 
the conventionally studied linear regime.

Let $C(t,t')=\langle A(t) B(t') \rangle$ be a 
correlation between observables $A$ and $B$, 
$R(t,t')=\delta \langle A(t) \rangle/\delta h(t') |_{h=0}$ the associated
response to a field conjugate to $B$,
and $\chi(t,t')\equiv \int_{t'}^t R(t,s) ds$ the integrated
response. The `differential' violation $V$ of   FDT at $(t,t')$ is
\begin{equation}
V (t,t') \equiv \frac{\partial C}{\partial t'} - TR= 
(1-X(t,t'))\frac{\partial C(t,t')}{\partial t'}
\; ,
\end{equation}
which also defines the `fluctuation-dissipation ratio' $X(t,t')$ \cite{Cuku}.  
Here, and in what follows, $t \geq t'$. A stronger and more physical form
of the violation is the integrated version $I(t,t') \equiv \int_{t'}^{t} V(t,s) ds$:
\begin{equation}
I(t,t')=C(t,t)-C(t,t')-T \chi(t,t')
\end{equation}
The static (zero-frequency) violation is
$I^{\omega=0}\equiv \lim_{t \to \infty} I(t,t')$.
We can now be more precise about what is meant by `far' from equilibrium.
 In the systems we consider,
$I(t,t')$ is (very) different from zero for certain large $t,t'$ 
even in the limit of vanishing entropy production
and, in particular, static susceptibilities do {\em not} coincide with 
their equilibrium value ($I^{\omega=0}\neq 0$)\cite{Cuku,review,foothoy,Frri}.

We assume a Langevin dynamics with an inertial term
\begin{equation}
m {\ddot x}_i + \gamma {\dot x}_i + \partial_{x_i}E +f_i =\Gamma_i,
\label{eq:ksde}
\end{equation} 
where $i=1,\dots,N$.
$f_i$ are velocity-independent non-conservative or time-dependent 
forces.
$\Gamma_i$ is a delta-correlated white noise  with variance 
$2\gamma T$.
This relation between the friction coefficient and the noise-correlation
is the `fluctuation -dissipation relation' (or FDT of the 
`second kind'); it expresses the fact that the bath itself
is and stays in equilibrium.
We encode $x_i$ in an $N$-vector ${\bbox x}$. 
For simplicity, we have set the Boltzmann constant $k_B=1$ and 
all masses $m_i$ to be equal to $m$. We  briefly describe the 
massless case at the end of the letter. 
We shall  not consider  purely Hamiltonian systems with  $\gamma =0$,
for reasons which will become evident.

The probability distribution at time $t$ for the process
(\ref{eq:ksde}) is given by 
$P({\bbox x},{\bbox v},t)= {\cal T} \exp( \int^t_o L_K) P({\bbox x},{\bbox v},0)$ 
with ${\cal T}$ denoting time-ordering and $L_K$ 
the Kramers operator given by
\cite{Risken}: 
\begin{equation}
L_K= -\partial_{x_i}v_i + \frac{1}{m} \partial_{v_i}
(\gamma v_i + \partial_{x_i}E +f_i + \gamma \frac{T}{m} \partial_{v_i})
\label{Kramers},
\end{equation}
where we have used the summation convention. 
An $H$-function  may be defined as \cite{Kubo,Mazur}
\begin{equation}
H(t)= \int d{\bbox x} d{\bbox v} P 
\left( T \ln P + E({\bbox x}) + \frac{m {\bbox v}^2}{2}  
\right)
\label{H}
\end{equation}
and may be  interpreted as a `generalized free-energy'. 

Using the equation for $P$ and  some integrations by parts one finds
\begin{equation}
{\dot H}(t)=-\langle {\bbox f} (t)  \cdot {\bbox v}(t) \rangle  -\sum_i g_i(t)
\; ,
\label{four}
\end{equation} 
where the first term is the power done by the forces ${\bbox f}$ and 
 $g_i$ are  the  entropy production terms  
\begin{equation}
g_i(t)=  \gamma \int d{\bbox x} d{\bbox v} 
\frac{(mv_i P+T \partial_{v_i}P)^2 }{m^2 P} \geq 0
\; .
\label{entr}
\end{equation}
Equations  (\ref{four}) and  (\ref{entr}) imply that in 
the purely relaxational ${\bbox f}=0$ case with bounded energy,
$H(t)$ is monotonically decreasing (and constant if  $\gamma=0$ --- Liouville's
theorem for Hamiltonian dynamics) and 
its time derivative must tend to zero since the equilibrium free energy 
is finite.
They  also  imply that a {\em stationary} (${\dot H}=0$)  driven system does
negative  external work on average.

We describe below a number of situations in which 
the differential FDT-violation $V$ vanishes
and the integral FDT-violation $I$, which can be finite, 
takes a restricted form:
\newline
(i) Purely relaxational systems \cite{review}
as the total entropy production rate tends to zero when $t' \to \infty$. 
This can be satisfied in 
{\em two ways}, depending on the large-times
sector we consider. 
For infinitely  separated times, `aging regime', one may have 
$X(t,t') \neq 1$ and $\partial_{t'} C \to 0$.
Instead, in the regime of  smaller time-separations
$X(t,t') \to 1$  and $\partial_{t'} C \neq 0$ (more like ordinary FDT).  
We also show that  the asymptotic rate of decay of $H(t)$ determines 
the  extent of both regimes.
\newline
(ii) Stationary driven systems  with ${\dot H}=0$ 
 in the limit of small driving power 
$\langle {\bbox f} \cdot {\bbox v} \rangle \to 0$.  
For time-separations that are larger for smaller driving powers 
one can have  \cite{driven}
 $V(t-t') \to 0$ with $X (t-t')\neq 1$ and $\partial_{t'}C(t-t') \to 0$,
while for time-differences that remain finite in the weak driving limit
$V(t-t') \to 0$ with  $X(t,t') \to 1$  and $\partial_{t'} C(t-t') \neq 0$.
\newline
(iii) Periodically driven systems of period $\tau$ that have achieved 
a stationary (period $\tau$) regime \cite{periodical} in the limit of vanishing 
work per cycle.  The FDT-violation over a cycle  
$\int_{t'}^{t'+\tau} ds V(t,s)$ vanishes with 
$
\int_{t'}^{t'+\tau} ds \langle {\bbox f(s)} \cdot {\bbox v(s)} \rangle
$.

In these three cases $X$ can be different from one precisely for pairs of times such
that the correlation evolves slowly. The fact that for large, 
widely separated times one can (and often does \cite{Cuku}) have $X \neq 1$ is 
crucial, because it ultimately leads to the violation of the integral
form of FDT: $I(t,t')\neq 0$ in certain large time sectors.

\noindent
(iv) Diffusion (with or without non-conservative forces) \cite{boge}
for times such that the root mean squared displacement at later time $t$ times 
the entropy production rate at the earlier time $t'$ vanishes.

\vspace{.2cm}
\noindent 
{\em Derivation.}

We  prove the result for macroscopic correlations and responses 
constructed as follows. 
Let $\{ A_i({\bbox x}) \}$ and $\{ B_i({\bbox x})\}$ be two sets 
of operators. We  assume that each $A_i$ ($B_i$) only depends  upon the
subset ${\cal C}^A_i$ (${\cal C}^B_i$) 
of the  degrees of freedom of the system.
We denote  $C_{A_i,B_i}(t,t')$ the correlation, $R_{A_i,B_i}(t,t')$ the response of
$A_i$ to a field conjugate to $B_i$ {\em in the energy}
and $V_{A_i,B_i}(t,t')\equiv  \partial_{t'}C_{A_i,B_i}-TR_{A_i,B_i}$
the corresponding FDT-violation. 
We  calculate the total FDT-violation 
$V_{AB}(t,t')$ associated with $N C_{AB} \equiv \sum_i^N C_{A_i,B_i}$ and 
$N R_{AB} \equiv \sum_i^N R_{A_i,B_i}$.

Using Eq.(\ref{Kramers}), one can easily show that
\begin{eqnarray}
V_{A_i,B_i}(t,t')
= \sum_l \theta(l,i)  \int d{\bbox x} d{\bbox v} d{\bbox x'} d{\bbox v'} A_i({\bbox x'}) 
\;\;\;\;\;\;\;\;\;\;
& &
\nonumber\\
\times 
 P({\bbox x'},{\bbox v'},t|{\bbox x},{\bbox v},t' ) 
\partial_{x_l} B_i({\bbox x}) 
\left(\frac{T}{m} \partial_{v_l} + v_l \right) P({\bbox x},{\bbox v},t')
& &
\label{ufa}
\end{eqnarray}
where 
$P({\bbox x'},{\bbox v'},t|{\bbox x},{\bbox v},t') \equiv \langle {\bbox x'},{\bbox v'} |{\cal T} e^{\int_{t'}^t L_K}|{\bbox x},{\bbox v}\rangle$.
We have made explicit the non-zero terms by introducing
 $\theta(l,i)=1$ if $l \in {\cal C}^B_i$ and zero otherwise. 

Identifying each term on the right  as 
$
\langle \Phi_{il}|\Psi_{il}  \rangle
\equiv 
\int d{\bbox x} d{\bbox v} d{\bbox x'} d{\bbox v'}
\Phi_{il} \Psi_{il}
$
with 
\begin{eqnarray}
\Phi_{il}
\equiv
A_i({\bbox x'})P^{1/2}({\bbox x'},{\bbox v'},t|{\bbox x},{\bbox v},t') P^{1/2}({\bbox x},{\bbox v},t')
\partial_{x_l} B_i({\bbox x})
\nonumber
\end{eqnarray}
and using the Cauchy-Scwhartz inequality 
$
|\langle \Phi_{il}|\Psi_{il} \rangle| 
\leq 
\sqrt{\langle \Phi_{il}|\Phi_{il} \rangle 
      \langle \Psi_{il}|\Psi_{il} \rangle  } 
$
one can bound separately each integral in (\ref{ufa}) as 
\begin{equation}
|\langle \Phi_{il}|\Psi_{il} \rangle| \leq 
\langle A_i^2(t) (\partial_{x_l}B_i)^2(t') \rangle ^{1/2}   
g_l^{1/2}(t') \theta(l,i)
\end{equation}
with $g_l(t')$ given by Eq.(\ref{entr}).  

The `macroscopic' FDT-violation is 
\begin{equation}
N V_{AB}(t,t') \equiv
\left|\sum_{i}
V_{A_i,B_i}(t,t')
\right| 
\; .
\end{equation}  
The Cauchy-Schwartz inequality 
applied on the sum yields
\begin{eqnarray}
N V_{AB}(t,t')
&\leq&  
\sum_{il} |\langle \Phi_{il}|\Psi_{il} \rangle|
\nonumber\\
&\leq &
\sqrt{N} D_{AB}(t,t') \left( 
\sum_{i'l'}
g_{l'}(t') \theta(l',i')
\right)^{1/2}
\label{ufa2}
\end{eqnarray}
where we have defined $D_{AB}$ through
\begin{eqnarray}
\sqrt{N} D_{AB}(t,t') \equiv 
\left(
\sum_{il}
\langle A_i^2(t) \left(\partial_{x_l}B_i\right)^2(t')  \rangle \theta(l,i)\right)^{1/2}
\; .
\nonumber
\end{eqnarray}
The last factor is $\sum_{l'} g_{l'}(t') \sum_{i'} \theta(l',i')=
\sum_{l'} g_{l'}(t') {\cal N}_{l'}$, where  ${\cal N}_{l'}$ is 
the number of different $A_{i'}$ that depend on  $x_{l'}$. 
Assuming all the  ${\cal N}_{l}$ to be finite,
 we can again bound the last factor using
${\cal N}_{l} \leq {\cal N}\equiv\max_{l}{\cal N}_{l}$. 
If in addition ${\cal C}_i^A$ has a finite number of elements,  $D_{AB}$ is $O(1)$. Thus
\begin{equation}
V_{AB}(t,t')
\leq \sqrt{{\cal N}} D_{AB}(t,t')
\left(\frac{1}{N} \sum_i g_i(t')
\right)^{1/2}  \; .
\label{bound}
\end{equation}
Both sides of the inequality are $O(1)$.
This is the basic result of this letter. We shall explore its consequences
 below.

For a purely relaxational system, this yields:
\begin{equation}
V_{AB}(t,t')
\leq \sqrt{{\cal N}} D_{AB}(t,t')
\left(-\frac{1}{\gamma N} \frac{dH (t')}{dt'}\right)^{1/2} 
\; .
\label{bound1}   
\end{equation}
Note that this proof breaks down in the non-dissipative limit $\gamma=0$. 

For stationary non-conservative systems we have
\begin{equation}
V_{AB}(t-t')
\leq \sqrt{{\cal N}} D_{AB}(t-t')
\left< \frac{{\bbox v} \cdot {\bbox f}}{N}\right>^{1/2} 
\; 
\label{bound2}
\end{equation}
while for a `stationary' periodically driven system
such that $H(t) = H(t+\tau)$:
\begin{eqnarray}
\int_{t'}^{t'+\tau} ds V_{AB}(t,s)
\leq \left({\cal N}   \int_{t'}^{t'+\tau}  ds D^2_{AB}(t,s) \right)^{1/2}
\left(\frac{W}{N}\right)^{1/2}
\nonumber
\end{eqnarray}
$W(t') \equiv \int_{t'}^{t'+\tau} ds \langle {\bbox v}(s) \cdot {\bbox f}(s)\rangle$
is the work per period.

These results can be generalized to 
establish a relation for
multiple point correlations between   operators  $A_i^k$ 
depending on the coordinates in 
${\cal C}_i$ at times $t_1 > ... > t_k$,
  $C_i^{1,\dots,k} \equiv \langle A_i^1 \dots A_i^k \rangle$,
and the corresponding responses $R_i^{1,\dots,k}$ to a perturbation applied 
to $A_i^k$ at time $t_k$. 

\vspace{.2cm}
\noindent
{\it Applications}

As a particular simple case of Eq.(\ref{bound1}) we obtain a bound 
for the variation of a single-time quantity in a relaxational case.
 Setting $A_i=1$ 
and taking a single operator $B$, $R_{AB}=0$  and we obtain
\begin{equation}
\left| \frac{d \langle B \rangle}{dt} \right|
\leq \left(
\sum_i^N \langle (\partial_{x_i} B)^2 \rangle\right)^{1/2}
\left(-\frac{1}{\gamma} \frac{dH(t)}{dt} \right)^{1/2}
\; .
\end{equation}

Another typical application of the formulae above is when 
$x_i$ are lattice variables, with $i$ denoting
the site. The fluctuation-dissipation theorem associated with the
two-point correlation $C(r,t,t')$ and response $R(r,t,t')$ is 
obtained by putting $B_i=x_i$ and $A_i=x_{i+r}$.
Similarly, one can study the energy-energy correlations 
$C^E(r,t,t')$ with $B_i=E_i-\langle E_i \rangle $ 
and $A_i=E_{i+r}-\langle E_{i+r} \rangle $,
 provided that
the energy of a site depends on a finite number of neighbours
 (and hence ${\cal N}$ is finite).

One can extend the derivation to the calculation of total correlations if the corresponding
spatial dependencies fall fast enough. For example, for the energy:
\begin{equation}
\frac{1}{N} \langle E(t)E(t') \rangle -\frac{1}{N^2}\langle E(t)
 \rangle \langle E(t') \rangle =  \sum_r  C^E(r,t,t')
\end{equation}
one can use the bounds obtained previously provided one can  cut off the sum at 
some maximum distance $r_{max}$.

Another interesting bound is obtained by writing 
$\langle {\bbox f}({\bbox x}) \cdot {\bbox v} \rangle =   
\int d{\bbox x} d{\bbox v} {\bbox f} ({\bbox x}) \cdot ( {\bbox v} + T \partial_{\bbox v})  P$
and by proceeding as from Eq.(\ref{ufa}) to Eq.(\ref{ufa2}):
\begin{equation}
\langle {\bbox f} \cdot {\bbox v} \rangle^2 
\leq
\langle | {\bbox f}|^2 \rangle \, \sum_i  g_i 
\; .
\end{equation}
 
\vspace{.2cm}
\noindent
{\em Integrated bounds for purely relaxational systems}

In glassy systems it is known \cite{Cuku,review,Frri} that the integral form of
 FDT is violated in certain two (large) time sectors.
We can bound the extent of these sectors as follows.

Assuming that the  factor $D_{AB}$ is finite  for all times, $D_{AB}(t,t')<K$,
on integrating   (\ref{bound1}) and applying once again
Cauchy-Schwartz, we obtain
\begin{eqnarray}
\left| I_{AB}(t,t') \right| 
\leq  K \int_{t'}^{t} \left( -\frac{1}{N}\frac{dH (s)}{ds}\right)^{1/2} ds
\; 
\nonumber
\end{eqnarray}
where $I_{AB}(t,t')=C_{AB}(t,t)-C_{AB}(t,t')-T\chi_{AB}(t,t')$.
Hence, there can be no integral violation
of FDT for {\em any} long times if $H $ falls faster than $t^{-1}$.
Interestingly enough, if
$H(t)$ can be written as an average over exponential 
processes
$H(t) = \int d\tau \rho(\tau) \exp(-t/\tau)$, 
this result implies that there can be FDT violation 
at large times only if  $\langle \tau \rangle$ diverges \cite{Bo}.

For $H $ falling slower than or as  $t^{-1}$, we can still bound the region in which FDT holds.
First of all, we have that for $t' \to \infty$ and $t-t'=O(1)$ FDT will always hold.
More generally, consider the limit $t' \to \infty$ with $t-t'=O({t'}^a)$  with  $0<a<1$. 
A simple calculation gives
\begin{eqnarray}
\left|
I_{AB}(t,t')
\right| 
\leq  K   \left(-\frac{1}{N}\frac{ dH (t')}{dt'}\right)^{1/2} \;  {t'}^a
\; .
\nonumber
\end{eqnarray}
If $H (t)= H _\infty + k t^{-\alpha}$, then there will be no FDT violation in the time sectors defined by
$a$ provided $2a<(\alpha+1)$. In particular, if $H $ falls as any inverse  power of a logarithm, 
FDT holds for $t-t'< O({t'}^{1/2})$.

\pagebreak

\noindent
{\em Massless case}

One can treat the case with $m=0$ ({\em i.e.} without inertial term) in the same way.
One defines an $H$ function as in (\ref{H}), without the kinetic term and with $P({\bbox x},t)$ depending only on space.
Using the Fokker-Planck equation, 
one may derive  an expression for the fluctuation-dissipation violation analogous to (\ref{ufa}), 
which can be bounded  in terms of the entropy production, which in this case reads:
\begin{equation}
g_i=   \int d{\bbox x}  \frac{(P \partial_{x_i}E({\bbox x})+
T \partial_{x_i}P )^2 }{P } \geq 0
\; .
\label{entr1}
\end{equation}

\vspace{.2cm}
\noindent
{\em Diffusion}

A simple application of the massless case is to
Brownian motion in $D$ dimensions, if we set the diffusion constant to
be equal to a temperature $T$, and
take $A_i=X_i$ and $B_i=X_i$
The application of the inequality for the massless case yields
\begin{eqnarray}
\left| V(t,t')\right| \leq (2DTt)^{1/2} 
\left(
T D/(2t')
\right)^{1/2} = T D\left(t/t'\right)^{1/2} 
\; . \nonumber 
\end{eqnarray}  
Note that here $D_{AB}(t,t') = 2Dt$ and is therefore
not bounded. Since in this case neither $R$ nor $\partial_{t'}C$ 
are small, this equation tells us that FDT can be violated \cite{Cukupa}. 
The exact calculation gives $V(t,t') = TD$, 
hence the inequality is clearly obeyed but with equality only at 
$t=t'$.

A more interesting example of the applications of our study is to the
problem of Sinai diffusion in one dimension \cite{diff,boge,Lale}. If one
assumes that ${\overline H(t)}$ scales in the same way as the energy 
${\overline E(t)}$, then one may deduce the scaling of the
energy via the Arrhenius law $t\sim \tau_0 \exp(c {\overline E(X_t)})$, with
$c$ and $\tau_0$ constants and $\overline{\;\; \cdot \;\;}$ representing an average 
over disorder. In this problem the particle is
subject to a white noise force and therefore is diffusing in a
Brownian potential, hence $\overline{E(x)} \sim x^{1/2}$. 
From this one deduces
the results $\langle X^2_t\rangle \sim \log^4(t) $ and $H(t)\sim
\overline{E(t)}\sim \log(t)$ (where we emphasize the 
relation between $\overline{E}$ and $\overline{H}$
is only justified on the grounds of physical intuition). We therefore
obtain $
\left| V(t,t')\right| \leq c \log^2(t)/\sqrt{t'}
$, which implies that the integrated form of FDT 
 must hold at least up to 
time differences scaling as $t-t' \sim c {t'}^{1/2}$.

In conclusion we have 
shown that there is a direct connection between FDT-violations and
entropy production in systems with stochastic dynamics. This connection allows
to obtain results even in the interesting cases in which in the limit
of small entropy production the systems are still far from equilibrium, as
may happen in  macroscopic (or diffusive) systems.
 It is important, though probably tougher,
to extend these results to deterministic systems with a 
thermostat \cite{thermo} in the thermodynamic limit;
it would be surprising if these results did not carry through to 
this case.

\vspace{.2cm}
\noindent
ACKNOWLEDGEMENTS. We wish to thank discussions with 
J-P Bouchaud,  G. Gallavotti, J-P Hansen,  P. Le Doussal, G. Lozano and 
D. Ruelle, and an anonymous referee.


\begin{references}
\bibitem{foothoy} See \cite{driven,Cukupe} for examples of stationary systems that violate FDT
in the limit of vanishing drive.
\end{references}
\end{document}